\documentclass[12pt,a4paper]{article}
\usepackage{a4wide}
\usepackage{epsfig}
\usepackage{cite}
\usepackage{amsmath}
\usepackage{graphicx}
\usepackage{latexsym}
\usepackage{array}

\newlength{\absize}
\newcommand{\dd}{\mbox{{\rm d}}}

\def\gsim{\mathrel{\rlap{\raise 2.5pt \hbox{$>$}}\lower 2.5pt
\hbox{$\sim$}}}
\def\lsim{\mathrel{\rlap{\raise 2.5pt \hbox{$<$}}\lower 2.5pt
\hbox{$\sim$}}}

\newcommand{\Lumint}{{\cal L}_{\rm int}}
\newcommand{\rpv}{{\not\!\! R_p}}

\usepackage[dvips]{color}
\definecolor{Black}{named}{Black}
\definecolor{Red}{named}{Red}

\begin{document}

\thispagestyle{empty}
\renewcommand{\thefootnote}{\fnsymbol{footnote}}
\newpage\normalsize
\pagestyle{plain}
\setlength{\baselineskip}{4ex}\par
\setcounter{footnote}{0}
\renewcommand{\thefootnote}{\arabic{footnote}}
\newcommand{\preprint}[1]{%
\begin{flushright}
\setlength{\baselineskip}{3ex} #1
\end{flushright}}
\renewcommand{\title}[1]{%
\begin{center}
\LARGE #1
\end{center}\par}
\renewcommand{\author}[1]{%
\vspace{2ex}
{\Large
\begin{center}
\setlength{\baselineskip}{3ex} #1 \par
\end{center}}}
\renewcommand{\thanks}[1]{\footnote{#1}}
\renewcommand{\abstract}[1]{%
\vspace{2ex}
\normalsize
\begin{center}
\centerline{\bf Abstract}\par
\vspace{2ex}
\parbox{\absize}{#1\setlength{\baselineskip}{2.5ex}\par}
\end{center}}

\vspace*{4mm} %\vfill

\title{Spin and model identification of $Z^\prime$ bosons at the LHC} \vfill

\author{P. Osland,$^{a,}$\footnote{E-mail: per.osland@ift.uib.no}
A. A. Pankov,$^{b,}$\footnote{E-mail: pankov@ictp.it} N.
Paver$^{c,}$\footnote{E-mail: nello.paver@ts.infn.it} and A. V.
Tsytrinov$^{b,}$\footnote{E-mail: tsytrin@rambler.ru}}

\begin{center}
$^{a}$Department of Physics and Technology, University of Bergen,
Postboks 7803, N-5020  Bergen, Norway\\
$^{b}$The Abdus Salam ICTP Affiliated Centre, Technical
University of Gomel, 246746 Gomel, Belarus\\
$^{c}$University of Trieste and INFN-Trieste Section, 34100
Trieste, Italy
\end{center}
%
%\keywords{Polarized $e^+e^-$ collisions, Models beyond the standard model}
%
%\pacs{12.60.-i, 12.60.Rc, 12.60.Cn}
%
\vfill

\abstract{Heavy resonances appearing in the clean Drell-Yan channel may be the
first new physics to be observed at the proton-proton CERN LHC. If a new
resonance is discovered at the LHC as a peak in the dilepton invariant mass
distribution, the characterization of its spin and couplings will proceed via
measuring production rates and angular distributions of the decay products.
We discuss the discrimination of the spin-1 of $Z^\prime$ representative
models ($Z^\prime_{\rm SSM}$, $Z^\prime_{\psi}$, $Z^\prime_{\eta}$,
$Z^\prime_{\chi}$, $Z^\prime_{\rm LR}$, and $Z^\prime_{\rm ALR}$) against the
Randall-Sundrum graviton resonance (spin-2) and a spin-0 resonance (sneutrino)
with the same mass and producing the same number of events under the observed
peak. To assess the range of the $Z^\prime$ mass where the spin determination
can be performed to a given confidence level, we focus on the angular
distributions of the Drell--Yan leptons, in particular we use as a basic
observable an angular-integrated center-edge asymmetry, $A_{\rm CE}$. The spin
of a heavy $Z'$ gauge boson can be established with $A_{\rm CE}$ up to
$M_{Z^\prime}\simeq 3.0$~TeV, for an integrated luminosity of 100 fb$^{-1}$,
or minimal number of events around 110.
We also examine the distinguishability of the considered $Z^\prime$ models
from one another, once the spin-1 has been established, using the total
dilepton production cross section. With some assumption, one might be able to
distinguish among these $Z^\prime$ models at 95\% C. L. up to
$M_{Z^\prime}\simeq 2.1$~TeV.}

\vspace*{20mm} \setcounter{footnote}{0} \vfill

\newpage
\setcounter{footnote}{0}
\renewcommand{\thefootnote}{\arabic{footnote}}

%\maketitle
%%%%%%%%%%%%%%%%%%%%%%%%%%%%%%%%%%%%%%%%%%%%%%%%%%%%%%%%%%%%%%%%%%%%%%%%%%%
\section{Introduction} \label{sect:introduction}
%%%%%%%%%%%%%%%%%%%%%%%%%%%%%%%%%%%%%%%%%%%%%%%%%%%%%%%%%%%%%%%%%%%%%%%%%%%
Heavy resonances with mass around 1~TeV or higher are predicted
by numerous New Physics (NP) scenarios, candidate solutions of
conceptual problems of the standard model (SM). In particular,
this is the case of models of gravity with extra spatial dimensions,
grand-unified theories (GUT), electroweak models with extended
spontaneously broken gauge symmetry, and supersymmetric (SUSY)
theories with $R$-parity breaking ($\rpv$). These new heavy
objects, or `resonances', with mass $M\gg M_{W,Z}$, may be
either produced or exchanged in reactions among SM particles at
high energy colliders such as the LHC and the
International electron-positron linear collider (ILC). A particularly
interesting process to be studied in this regard at the LHC
is the Drell-Yan (DY) dilepton production ($l=e, \mu$)
\begin{equation}
p+p\to l^+l^-+X, \label{proc_DY}
\end{equation}
where exchanges of the new particles can occur and manifect themselves as
peaks in the $(l^+l^-)$ invariant mass $M$. Once the heavy resonance is
discovered at some $M=M_R$, further analysis is needed to identify the
theoretical framework for NP to which it belongs. Correspondingly, for any NP
model, one defines as {\it identification} reach the upper limit for the
resonance mass range where it can be identified as the source of the
resonance, against the other, potentially competitor scenarios, that can give
a peak with the same mass and same number of events under the peak. This
should be compared to the {\it discovery} reach, which specifies the
(naturally more extended) mass range where the peak in the cross section
pertaining to the model can just be observed experimentally. Clearly, the
determination of the spin of the resonance represents an important aspect of
the selection among different classes of non-standard interactions giving rise
to the observed peak.  
\par
Tests of the spin-2 of the Randall-Sundrum~\cite{Randall:1999ee} graviton
excitation (RS) exchange in the process (\ref{proc_DY}) at LHC, against the
spin-1 hypothesis, have been recently performed, e.g., in
Refs.~\cite{Allanach:2000nr,Allanach:2002gn,Cousins:2005pq} on the basis of
the lepton differential polar angle distribution, and in
Ref.~\cite{Murayama:2009jz} using the azimuthal angular dependence.  In the
reverse, the identification of the spin-1 $Z^\prime$s has been discussed in
\cite{Feldman:2006wb,Petriello:2008zr}. The above-mentioned differential
angular analysis in the polar angle has been applied to the search for spin-2,
spin-1 and spin-0 exchanges in the experimental studies of process
(\ref{proc_DY}) at the Fermilab Tevatron proton-antiproton collider
\cite{Abulencia:2005nf}.

\par In
Ref.~\cite{Osland:2008sy}, the discrimination reach at the LHC on the spin-2
RS graviton resonance or, more precisely, the simultaneous rejection of {\it
both} the spin-1 and spin-0 hypotheses for the peak, has been assessed by
using as basic observable an angular-integrated center-edge asymmetry, $A_{\rm
CE}$, instead of the `absolute' lepton differential angular distribution. The
potential advantages of the asymmetry $A_{\rm CE}$ to discriminate the spin-2
graviton resonance against the spin-1 hypothesis were discussed in
Refs.~\cite{Osland:2003fn,Dvergsnes:2004tw}.  

\par Here, along the lines of Ref.~\cite{Osland:2008sy} but in the reverse
direction, we apply the same basic observable $A_{\rm CE}$, to the spin-1
identification of a peak observed in the dilepton mass distribution of process
(\ref{proc_DY}) at the LHC, against the spin-2 and spin-0 alternative
hypotheses. For explicit NP realizations, for the spin-1 $Z^\prime$ models we
refer to Refs.~\cite{Hewett:1988xc}; for the alternative spin-2 and spin-0
hypotheses we refer for the RS graviton resonance to~\cite{Randall:1999ee} and
for the SUSY $\rpv$ sneutrino exchange to
\cite{Kalinowski:1997bc,Allanach:1999bf}, respectively.

\par It turns out that $A_{\rm CE}$ should provide a robust
spin diagnostic for the spin-1 case also. Moreover, we examine the
possibility, once the spin-1 for the discovered peak is established, of
differentiating the various representative $Z^\prime$ models from one
another. For this purpose, we must use the total dilepton production cross
section or, equivalently, the rate of events of reaction (\ref{proc_DY}) under
the peak.  Identification of $Z^\prime$ models have been discussed recently in,
e.g.~\cite{Dittmar:2003ir,Cousins:2005uf,Petriello:2008zr} with different sets
of observables, namely, forward-backward asymmetry $A_\text{FB}$ on and off
the $Z^\prime$ resonance, $Z^\prime$ rapidity distribution, cross section
times total width, $\sigma\times\Gamma_{Z^\prime}$,
as well as in different processes \cite{Godfrey:2005pm,Godfrey:2008vf}.
It was found that, on the basis of $A_\text{FB}$ 
only, pairs of $Z^\prime$ models become indistinguishable at a 
given level of significance, starting from relatively low values of 
$M_{Z^\prime}$ of the order of 1--2 TeV, even at $\Lumint$ 
much higher than $100~\text{fb}^{-1}$. 
These ambiguities can be reduced by the combined analysis of the 
observables mentioned above, and at $\Lumint=100~\text{fb}^{-1}$, 
some models could be discriminated up to $Z^\prime$ mass
of the order of 2--2.5 TeV. As we will note below, on the 
basis of a simple $\chi^2$ criterion, the precise determination 
of the total cross section itself might provide  
a somewhat stronger discrimination potential, in the sense that 
all models could be pairwise distinguished from one another up 
to $Z^\prime$ masses of about 2.1 TeV. 

\par In Sec.~\ref{sect:models} we present a brief introduction to the main
features of the different models considered in the analysis, and the expected
relevant statistics; Sec.~\ref{sect:spin1-ID} is devoted to the spin-1
identification of $Z^\prime$ bosons against the spin-2 RS and the spin-0
sneutrino hypotheses; in Sec.~\ref{sect:LHCdiff} we derive the differentiation
of $Z^\prime$ models among themselves obtainable at the LHC from consideration
of total dilepton cross sections, wheras Sec.~\ref{sect:low-low} is devoted to
a brief discussion of the reduced-energy, low-luminosity domain relevant to
the early running period of the collider.  Finally, Sec.~\ref{sect:Concl}
contains some conclusive remarks.
%%%%%%%%%%%%%%%%%%%%%%%%%%%%%%%%%%%%%%%%%%%%%%%%%%%%
\section{Cross sections and considered NP models}
\label{sect:models}
\setcounter{equation}{0}
%%%%%%%%%%%%%%%%%%%%%%%%%%%%%%%%%%%%%%%%%%%%%%%%%%%
For completeness and to fix the notations, we start by recalling the basic
expression for the cross section of process (\ref{proc_DY}), and present a
mini-review of the NP models we want to compare.
\par The parton model cross section for inclusive production of a dilepton
with invariant mass $M$ can be written as
\begin{equation}
\label{Eq:dsigma-dMdydz}
\frac{{\rm d}\sigma(R_{ll})}{{\rm d} M\, {\rm d} y\, {\rm d} z}
= K\frac{2M}{s}
\sum_{ij} f_i(\xi_1,M)f_j(\xi_2,M)
\frac{{\rm d}\hat\sigma}{{\rm d}z} (i+j\to l^++l^-).
\end{equation}
Here, $s$ is the proton-proton center-of-mass energy
squared; $z=\cos\theta_{\rm c.m.}$ with $\theta_{\rm c.m.}$
the lepton-quark angle in the dilepton center-of-mass
frame; $y$ is the dilepton rapidity; $f_{i,j}(\xi_{1,2},M)$
are parton distribution functions in the protons $P_1$ and
$P_2$, respectively, with
$\xi_{1,2}=(M/\sqrt s)\exp(\pm y)$ the parton
fractional momenta; finally, ${\rm d}\hat\sigma_{ij}$ are the
partonic differential cross sections.
In~(\ref{Eq:dsigma-dMdydz}),
the factor $K$ accounts for next-to-leading order QCD
contributions~\cite{Carena:2004xs,Mathews:2004xp}. For
simplicity, and to make our procedure more transparent,
we will use as an approximation a global flat value $K=1.3$.
\par
Since we are interested in a (narrow) peak production and
subsequent decay into the DY pair, $pp\to R\to l^+l^-$, we
consider the lepton differential angular distribution,
integrated over an interval of $M$ around $M_R$:
\begin{equation}
\frac{\dd\sigma(R_{ll})}{\dd z}
=\int_{M_{R}-\Delta M/2}^{M_{R}+\Delta M/2}\dd M
\int_{-Y}^{Y}\frac{\dd\sigma}{\dd M\, \dd y\, \dd z}\,\dd y.
\label{Eq:DiffCr}
\end{equation}
The number of events under the peak, that determines the
statistics, is therefore given by:
\begin{equation}
\sigma(R_{ll})\equiv\sigma{(pp\to R)} \cdot
\text{BR}(R \to l^+l^-)
=\int_{-z_{\text{cut}}}^{z_\text{cut}}\dd z
\int_{M_{R}-\Delta M/2}^{M_{R}+\Delta M/2}\dd M
\int_{-Y}^{Y}\dd y
\frac{\dd\sigma}{\dd M\, \dd y\, \dd z}.
\label{Eq:TotCr}
\end{equation}
For the full final phase space, $z_{\rm cut}=1$ and
$Y=\log(\sqrt{s}/M)$. However, if the finite detector angular
acceptance is accounted for, $z_{\rm cut}<1$ and $Y$ in
Eqs.~(\ref{Eq:DiffCr}) and (\ref{Eq:TotCr}) must be replaced by a
maximum value $y_{\rm max}(z,M)$. Concerning the size of the
bin $\Delta M$, it should include a number (at least one)
of peak widths to enhance the probability to pick up the
resonance. In the models we will consider, widths are
predicted to be small, typically of the order of
a percent (or less) of the mass $M_R$, so that the
integral under the peak should practically be
insensitive to the actual value of $\Delta M$.
Conversely, the SM `background' is expected
to depend on $\Delta M$.
In our analysis, we adopt the parametrization of
$\Delta M$ vs.\ $M$ proposed in Ref.~\cite{Feldman:2006wb}
and, denoting by $N_B$ and $N_S$ the number of
`background' and `signal' events in the bin, the
criterion $N_S=5{\sqrt{N_B}}$ or 10~events,
whichever is larger,
as the minimum signal for the peak discovery.
\par To evaluate the statistics, we shall use in
Eqs.~(\ref{Eq:DiffCr}) and (\ref{Eq:TotCr})
the CTEQ6.5 parton
distributions~\cite{Pumplin:2002vw}, and impose cuts
relevant to the LHC detectors, namely: pseudorapidity
$\vert\eta\vert<2.5$ for both leptons assumed massless
(this leads to a  boost-dependent cut on
$z$~\cite{Dvergsnes:2004tw}); lepton transverse momentum
$p_\perp > 20\, {\rm GeV}$. Moreover, the reconstruction
efficiency is taken to be 90\% for both electrons and muons
\cite{Cousins:2004jc} and throughout this paper, except for
Sec.~\ref{sect:low-low}, a time-integrated LHC luminosity
${\cal L}_{\rm int}=100\, {\rm fb}^{-1}$.
\par
For the proton-proton initiated process (\ref{proc_DY}),
only the $z$-even parts of the partonic differential
cross sections contribute to the right-side of Eq.~(\ref{Eq:DiffCr}),
$z$-odd terms do not contribute after
the $y$-integration.\footnote{Accordingly, for
the $q\bar q$ and $gg$ subprocesses, only the
combinations of parton distributions
$[f_q(\xi_1,M)f_{\bar q}(\xi_2,M)+f_{\bar q}(\xi_1,M)f_q(\xi_2,M)]$
and $f_g(\xi_1,M)f_g(\xi_2,M)$ are effective in the cross
sections (\ref{Eq:DiffCr}) and (\ref{Eq:TotCr}).}
Also, due to $M_Z\ll M_R$ and the
narrow width peak, the resonant amplitude interference with
the SM is expected to give negligible contributions to the
right-hand sides of (\ref{Eq:DiffCr}) and (\ref{Eq:TotCr})
after the symmetric $M$-integration around $M_R$
needed there. Thus, we can retain in these equations
just the SM and the resonance pole
contributions.\footnote{Actually, such interference can
in principle contribute appreciably to the differential
cross section $\dd\sigma/\dd M \dd y$ \cite{Osland:2008sy}, 
and plays a role in the
forward--backward asymmetry (which we do not consider here).} 
This fact was noticed for $Z^\prime$-exchange in, e.g.,
Refs.~\cite{Feldman:2006wb, Carena:2004xs}, but holds
also for the RS graviton and for
the scalar sneutrino exchanges discussed later.

%%%%%%%%%%%%%%%%%%%%%%%%%%%%%%%
\subsection{$Z^\prime$ models}
\label{subsect:Zprime}
%%%%%%%%%%%%%%%%%%%%%%%%%%%%%%
In a wide variety of electroweak theories, in particular those
based on extended, spontaneously broken, gauge symmetries, the
existence of one (or more) new neutral gauge bosons $Z^\prime$ is
envisaged. These additional gauge bosons could be accessible at
the LHC.  A new neutral gauge boson would induce additional
neutral current interactions.
The color-averaged differential cross section for the relevant,
leading order, partonic subprocess {${q{\bar q}\to Z^\prime\to
l^+l^-}$} can be expressed as:\footnote{We neglect fermion masses
as well as potential effects from the (tiny) $Z-Z^\prime$ mixing.}
\begin{equation}
\frac{\dd\hat \sigma_{q\bar q}^{Z^\prime}} {\dd
z}\bigg\vert_{z{\rm -even}}=\frac{1}{N_c}\, \frac{\pi\alpha_{\rm
em}^2}{2 M^2}\,[S_q^{Z^\prime}\, (1+z^{2})], \label{partZprime}
\end{equation}
with
\begin{equation}
\label{Eq:zprime-partonlevel} S_q^{Z^\prime}=\frac{1}{4}\,
({{{g_L^q}^\prime}}^2+{{{g_R^q}^\prime}}^2)\,({{{g_L^l}^\prime}}^2+
{{{g_R^l}^\prime}}^2)\vert\chi_{Z^\prime}\vert^2, 
\qquad 
\chi_{Z^\prime}=\frac{M^2}{{M^2 - M_{Z^\prime}^2 
+ i\,M_{Z^\prime} \Gamma _{Z^\prime} }}.
\end{equation}
Eq.~(\ref{partZprime}) shows that the spin-1, $Z^\prime$, exchange in process
(\ref{proc_DY}) has the same symmetrical angular dependence as the SM $\gamma$
and $Z$ exchanges.  
\par The list of $Z^\prime$ models that will be considered in our analysis is
the following:
\begin{itemize}
\item[(i)]
The three possible $U(1)$ $Z'$ scenarios originating from
the exceptional group $E_{6}$ spontaneous breaking. 
They are defined in terms of a
mixing angle $\beta$, and the couplings are as in Tab.~\ref{Table:couplings}. 
The specific values $\beta=0$, $\beta=\pi/2$
and $\beta=\arctan{-\sqrt{5/3}}$, correspond to different $E_{6}$
breaking patterns and define the popular scenarios $Z^\prime_\chi$,
$Z^\prime_\psi$ and $Z^\prime_\eta$, respectively.

\item[(ii)] The left-right models, originating from the breaking of an
$SO(10)$ grand-unification symmetry, and where the corresponding
$Z^\prime_{\rm LR}$ couples to a combination of right-handed and $B-L$
neutral currents ($B$ and $L$ denote lepton and baryon currents), specified by
a real parameter $\alpha_{\rm LR}$ bounded by $\sqrt{2/3} \lsim \alpha_{\rm
LR}\lsim\sqrt{2}$. Corresponding $Z^\prime$ couplings are reported in
Tab.~\ref{Table:couplings}. We fix $\alpha_{\rm LR}=\sqrt 2$, which
corresponds to a pure L-R symmetric model.

\item[(iii)]
The $Z'_{\rm ALR}$ predicted by the `alternative' left-right scenario.

\item[(iv)]
The so-called sequential $Z^\prime_\text{SSM}$, where the
couplings to fermions are the same as those of the SM $Z$.
\end{itemize}
Detailed descriptions of these models, as well as the specific
references, can be found, e.~g., in Ref.~\cite{Hewett:1988xc}. All
numerical values of the $Z^\prime$ couplings needed in
Eq.~(\ref{Eq:zprime-partonlevel}) are collected in
Table~\ref{Table:couplings}, where:
$A=\cos\beta/2\sqrt{6}$ and
$B=\sqrt{10}\sin\beta/12$ are used. 

Current $Z^\prime$ mass limits, from the Fermilab Tevatron
collider, are in the range $800-1000~\text{GeV}$, depending on the
model~\cite{Tev:2007sb}.

%------------------------------
\begin{table}[htb]
\caption{\label{Table:couplings} Left- and right-handed
couplings of the first generation of SM fermions to the $Z^\prime$
gauge bosons, in units of $1/c_W$ for the $E_6$ and LR models,
and $1/s_Wc_W\sqrt{1-2s_W^2}$ for the ALR model \cite{Osland:2008sy},
where $c_W=\cos\theta_W$, $s_W=\sin\theta_W$.}
\begin{center}
\begin{tabular}{|c|c|c|c|c|} \hline
\multicolumn{5}{|c|}{$E_6$ model}\\ \hline \hline
fermions ($f$)  & $\nu$ & $e$ & $u$ & $d$ \\
\hline %\hline
$g_L^{f\prime}$ & $3A+B$ & $3A+B$ & $-A+B$ & $-A+B$ \\
\hline
$g_R^{f\prime}$ & 0 & $A-B$ & $A-B$ & $-3A-B$ \\
\hline \hline \multicolumn{5}{|c|}{Left-Right model (LR)}\\
\hline \hline $g_L^{f\prime}$ &
${\frac{1}{2\,\alpha_{LR}}}$ & ${\frac{1}{2\,\alpha_{LR}}}$ &
$-{\frac{1}{ 6\,\alpha_{LR}}}$ &
$-{\frac{1}{6\,\alpha_{LR}}}$ \\
\hline $g_R^{f\prime}$ & 0 &
${\frac{1}{2\,\alpha_{LR}}}- {\frac{\alpha_{LR}}{ 2}}$ &
$-{\frac{1}{6\,\alpha_{LR}}}+{\frac{\alpha_{LR}}{2}}$ &
$-{\frac{1}{6\,\alpha_{LR}}}-{\frac{\alpha_{LR}}{2}}$ \\
\hline \hline \multicolumn{5}{|c|}{Alternative Left-Right model (ALR)}\\
\hline \hline $g_L^{f\prime}$ &
${-\frac{1}{2}+s_W^2}$ & ${-\frac{1}{2}+s_W^2}$ &
${-\frac{1}{6}s_W^2}$ &
${-\frac{1}{6}s_W^2}$ \\
\hline $g_R^{f\prime}$ & 0 &
${-\frac{1}{2}+\frac{3}{2}s_W^2}$ &
${\frac{1}{ 2}-\frac{7}{6}s_W^2}$ & $\frac{1}{3}s_W^2$ \\
\hline
\end{tabular}
\end{center}
\end{table}

%%%%%%%%%%%%%%%%%%%%%%%%%%%%%%%%%%%%%%%%%%%%%%%%%%%%%%%%%%%
%%%%%%%%%%%%%%%%%%%%%%%%%%%%%%%%%%%%%%%%%%%%%%%%%%%%%%%%%%%
\subsection{RS graviton excitation}
\label{subsect:graviton}
%%%%%%%%%%%%%%%%%%%%%%%%%%%%%%%%%%%%%%%%%%%%%%%%%%%%%%%%%%
We consider the simplest scenario in the class of
models based on one compactified warped extra
dimension and two branes, proposed in the context of 
the SM gauge-hierarchy problem in~\cite{Randall:1999ee}. 
The model predicts a tower of narrow Kaluza--Klein (KK), 
spin-2, graviton excitations $G^{(n)}$ ($n\ge 1$) 
with the peculiar mass spectrum 
$M^{(n)}=M^{(1)} x_n/x_1$ ($x_i$ are the zeros
of the Bessel function, $J_1(x_i)=0$). Their masses and
couplings to the SM particles are proportional to
$\Lambda_\pi$ and $1/\Lambda_\pi$, respectively, with
$\Lambda_\pi$ the gravity effective mass scale on the
SM brane. For $\Lambda_\pi$ of the TeV order, such RS 
graviton resonances can be exchanged in the process 
(\ref{proc_DY}) and mimic $Z^\prime$ exchange. 
The independent parameters of the model 
can be chosen as the dimensionless ratio 
$c=k/{\overline M}_{\rm Pl}$ (with $k$ the 5-dimensional 
curvature and ${\overline M}_{\rm Pl}=1/\sqrt{8\pi G_{\rm N}}$ 
the reduced Planck mass), and the mass $M_G$ of the lowest KK 
resonance $G^{(1)}$. Accordingly, $\Lambda_\pi=M_G/cx_1$.

\par
The differential cross sections for the relevant partonic
subprocesses needed in (\ref{Eq:DiffCr}) and (\ref{Eq:TotCr}),
$q\bar q\to G\to l^+l^-$ and $gg\to G\to l^+l^-$, read,
with $\kappa=\sqrt{2}cx_1/M_G$
\cite{Allanach:2000nr,Han:1998sg,Giudice:1998ck,Bijnens:2001gh,
Dvergsnes:2002nc}
\begin{equation}
\frac{{\dd\hat{\sigma}_{q\bar q}^G }}{{\dd z}}
+ \frac{{\dd\hat{\sigma}_{gg}^G }}{{\dd z}}\bigg\vert_{z{\ \rm even}}
= \frac{{\kappa ^4 M^2 }}{{640\pi ^2 }}\left[ {\Delta _{q\bar q} (z)
+ \Delta_{gg} (z)} \right]\vert\chi_G\vert^2;
\quad \chi_G=\frac{M^2}{{M^2 - M_G^2 + i\,M_G \Gamma _G }}
\label{partG}
\end{equation}
where
\begin{equation}
\Delta_{q\bar q} (z)
= \frac{\pi }{{8N_C }}\,\frac{5}{8}\,(1 - 3z^2  + 4z^4 ),
\hspace{1cm}
\Delta _{g g} (z) = \frac{\pi }{{2(N_C^2  - 1)}}\,\frac{5}{8}\,(1 - z^4).
\label{angfuncs}
\end{equation}
The theoretically `natural' ranges for the RS model
parameters are $0.01\leq c\leq 0.1$ and
$\Lambda_\pi<10\, {\rm TeV}$~\cite{Davoudiasl:2000jd}.
Current lower bounds at 95\% C.L. from the Fermilab
Tevatron collider are, for the first graviton mass:
$M_G>300$ GeV for $c=0.01$ and $M_G>900$ GeV for
$c=0.1$~\cite{Tev:2007sb,Abazov:2007ra}.

%%%%%%%%%%%%%%%%%%%%%%%%%%%%%%%%%%%%%%%%%%%%%%%%%
\subsection{Sneutrino exchange}
\label{subsect:sneutrino}
%%%%%%%%%%%%%%%%%%%%%%%%%%%%%%%%%%%%%%%%%%%%%%%%%
Sneutrino ($\tilde\nu$) exchange can occur in SUSY with
$R$-parity breaking, and represents a possible, spin-0,
interpretation of a peak in the
dilepton invariant mass distribution of the process
(\ref{proc_DY}). The cross section for the relevant
partonic process, $q\bar{q} \to\tilde{\nu} \to l^+l^-$,
is flat in $z$ and reads~\cite{Kalinowski:1997bc}:
\begin{equation}
\frac{\dd\hat{\sigma}_{q\bar q}^{\tilde{\nu}}}{{\dd z}}
= \frac{1}{N_c}\,\frac{\pi\alpha_\text{em}^2}{4\,M^2}
\left(\frac{\lambda\lambda^\prime}{e^2}\right)^2
\vert\chi_{\tilde\nu}\vert^2\,\delta_{qd},
\qquad
\chi_{\tilde\nu}=\frac{M^2}{{M^2 - M_{\tilde\nu}^2 +
i\, M_{\tilde\nu} \Gamma _{\tilde\nu} }}.
\label{cross_snu}
\end{equation}
In Eq.~(\ref{cross_snu}), $\lambda$ and $\lambda^\prime$
are the $R$-parity-violating sneutrino couplings to
$l^+l^-$ and $d{\bar d}$, respectively. Actually, in
the narrow-width approximation, the cross section (\ref{cross_snu})
turns out to depend on the product
$X=(\lambda^\prime)^2B_l$, with $B_l$ the
sneutrino leptonic branching ratio. Current limits
on $X$ are rather loose~\cite{constraints}, and we
may consider for this parameter the range
$10^{-5} \leq X \leq 10^{-1}$.
For $10^{-4} \leq X \leq 10^{-2}$, the range is 
$M_{\tilde\nu}\gsim280-800~\text{GeV}$ \cite{Tev:2007sb}.
%%%%%%%%%%%%%%%%%%%%%%%%%%%%%%%%%%%%%%%%%%%%%%%%%%%%
\begin{figure}[tbh!] % Fig.1
\vspace*{-0.5cm}
\centerline{ \hspace*{-2.0cm}
\includegraphics[width=12.0cm,angle=0]{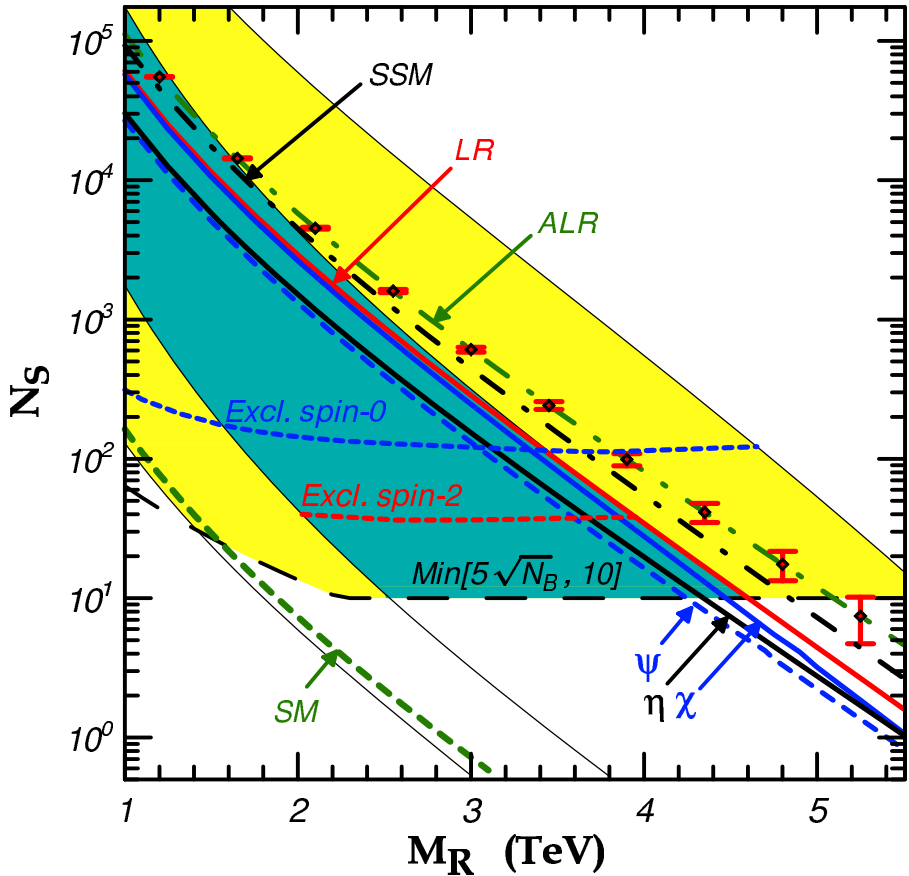}}
%\vspace*{-7.5cm}
\caption{\label{fig1} Expected number of resonance (signal) events $N_S$ vs.\
$M_{R}$ ($R=Z^\prime,G,\tilde\nu_\tau$) at the LHC with
$\Lumint=100~\text{fb}^{-1}$ for the process $pp\to R\to l^+l^-+X$
($l=e,\mu$). Event rates for various $Z^\prime$ models are shown. Green area
corresponds to graviton signature space for $0.01<c<0.1$ while the yellow area
is the sneutrino signature space for $10^{-5}<X<10^{-1}$. Minimum number of
signal events needed to detect the resonance (5-$\sigma$ level) above the
background and the minimum number of events to exclude the spin-2 and spin-0
hypotheses at 95\% C.L. are shown. Error bars correspond
to the statistical uncertainties for the ALR model.}
\end{figure}
%%%%%%%%%%%%%%%%%%%%%%%%%%%%%%%
\subsection{Statistics and model signature spaces}
\label{subsect:statistics}
%%%%%%%%%%%%%%%%%%%%%%%%%%%%%%%%%%%%%%%%%%
In Fig.~\ref{fig1}, %\ref{fig1}
we show the predicted number of resonance (signal)
events $N_S$ in the Drell-Yan process (\ref{proc_DY})
at LHC, {\it vs.}\ $M_{R}$, where $R=Z',G,\tilde\nu$
denotes the three alternative possibilities outlined
in the previous subsections. The assumed integrated
luminosity is ${\cal L}_{\rm int}=100\,\text{fb}^{-1}$,
the cuts in phase space relevant to the
foreseen detector acceptance specified above have
been imposed, and the channels $l=e,\mu$ have been
combined. Also, the minimum
signal for resonance discovery above the `background'
at 5$\sigma$ is represented by the long-dashed line.
\par
For any model, one can define a corresponding
{\it signature space} as the region, in the ($M_R,N_S$)
plot of Fig.~\ref{fig1}, that can be `populated' by the model
by varying its parameters in the domains mentioned
above. Clearly, in  regions where the signature
spaces overlap, the values of $M_R$ are such that
it is not possible to distinguish a model as the
source of the peak against the others, because the
number of signal events under the peak can be
the same. Further analyses are needed in these cases
to perform the identification of the peak source.
For example, the `blue' area in Fig.~\ref{fig1} corresponds
to the graviton signature space for
$0.01\leq c\leq 0.1$, while the yellow area
(which has substantial overlap with the blue
one---indicated as green) is that for the sneutrino signature space
corresponding to $10^{-5}\leq X\leq 10^{-1}$.
\par
As regards the discovery and identification
of $Z^\prime$ we are interested in, the signature
spaces in Fig.~\ref{fig1} reduce to the lines labelled by
the different models, because the event rates
are fixed, once $M_Z^\prime$ is given, through
the couplings in Table~1. Fig.~\ref{fig1} shows that,
with the assumed (high) luminosity of
$100\, {\rm fb}^{-1}$, $Z^\prime$ gauge boson
masses up to 4--5 TeV are in principle within the
5-$\sigma$ reach of the LHC, consistent with earlier
studies \cite{Hewett:1988xc}.
We here assume that the $Z'$ can
only decay to pairs of SM fermions in order to
obtain the leptonic branching ratio $B_l$. It is
important to note that in many models, where $Z'$
can also decay to exotic fermions and/or SUSY
particles this overestimates $B_l$ and, thus, the
search reach.
\par On the other hand, Fig.~\ref{fig1} demonstrates that,
as far as the production rate of DY pairs is concerned,
there is a substantial overlap between the $Z^\prime$
and the $\tilde\nu$ signature spaces, which determines
a domain in ($M_{\tilde\nu},X$) where spin-0 $\tilde\nu$
exchange and $Z^\prime$ exchanges are not distinguishable
because they lead to the same event rate under the peak.
The same is true for the spin-2, RS model. However,
as shown by Fig.~\ref{fig1}, in this case it is interesting
to note that, if one literally takes the suggested
range $c\leq 0.1$ as the `naturally' preferred one,
the ALR and SSM scenarios can be discriminated
against the RS (spin-2) resonance {\it already} at
the level of event rates in a wide range of mass values
accessible to the LHC, with no need for further analyses.
Conversely, only the $E_6$ and LR $Z^\prime$ models
possess a `confusion region' with the RS resonance $G$,
concentrated near the upper border of the graviton
signature domain.
%%%%%%%%%%%%%%%%%%%%%%%%%%%%%%%%%%%%%%%
\section{Identification of $Z^\prime$ spin-1}
\label{sect:spin1-ID}
\setcounter{equation}{0}
%%%%%%%%%%%%%%%%%%%%%%%%%%%%%%%%%%%%%%%
We now turn to the identification of the spin-1
of the $Z^\prime$ boson {\it vs.} the spin-0 and spin-2
hypotheses using the angular distribution of the
final-state leptons.

\par
For this purpose, we adopt the integrated center-edge
asymmetry $A_{\rm CE}$, defined as \cite{Osland:2003fn,Dvergsnes:2004tw}:
\begin{equation}
\label{Eq:ace}
A_{\rm{CE}}(M_R)=\frac{\sigma_{\rm{CE}}(R_{ll})}
{\sigma(R_{ll})},\quad {\rm with}\quad
\sigma_{\rm{CE}}(R_{ll}) \equiv
\left[\int_{-z^*}^{z^*} - \left(\int_{-z_\text{cut}}^{-z^*}
+\int_{z^*}^{z_\text{cut}}\right)\right] \frac{\dd
\sigma(R_{ll})}{\dd z}\, \dd z.
\end{equation}
In Eq.~(\ref{Eq:ace}), $R$ denotes the three hypotheses
for the resonances we want to compare, namely: spin-1 ($V$);
spin-2 ($G$); and spin-0 ($S$). Moreover,
$0<z^*<z_{\rm cut}$ is an a priori free value of
$\cos\theta_{\rm cm}$ that defines the `center' and `edge'
angular regions.

\par
Using the differential partonic cross sections reported
in the previous sections, one finds the explicit
$z^*$-dependencies of $A_{\rm CE}$ for the three cases:
\begin{align}
\label{ACE1} A_{\rm CE}^{V} \equiv A_{\rm CE}^{\text{\rm SM}}
&=\frac{1}{2}\,z^*({z^*}^2+3)-1, \\
\label{ACE2}
A_{\rm CE}^{G} &=\epsilon_q^{\rm SM}\,A_{\rm CE}^{V} +
\epsilon_q^G\left[2\,{z^*}^5+\frac{5}{2}\,z^*(1-{z^*}^2)-1\right]
\nonumber \\
&+ \epsilon_g^G\left[\frac{1}{2}\,{z^*}(5-{z^*}^4)-1\right], \\
\label{ACE0} A_{\rm CE}^{S} &= \epsilon_q^{\rm SM}\,A_{\rm CE}^{V}
+\epsilon_q^{S}\,(2\,z^*-1).
\end{align}
In Eq.~(\ref{ACE2}), $\epsilon_q^G$, $\epsilon_g^G$ and
$\epsilon_q^{\rm SM}$ are the fractions of resonant events from
the processes $\bar qq,gg\to G\to l^+l^-$ and from the SM
background, respectively, with $\epsilon_q^G+\epsilon_g^G+
\epsilon_q^{\rm SM}=1$. Analogous definitions hold for
Eq.~(\ref{ACE0}), where now $\epsilon_q^S+ \epsilon_q^{\rm SM}=1$.
Their dependence on the dilepton invariant mass $M$ is determined
by the overlap of parton distribution functions in
Eqs.~(\ref{Eq:dsigma-dMdydz}) and (\ref{Eq:DiffCr}). Actually, the
above equations strictly hold for $z_{\rm cut}=1$, while all the
results and figures reported here will be obtained by taking
detector cuts into account. Differences turn out to be
appreciable, and have an impact on the assessment of
identification reaches, only near $z^*=1$, whereas the `optimal'
values used in the $A_{\rm CE}$ analysis will be $z^*\approx 0.5$.
Thus, Eqs.~(\ref{ACE1})--(\ref{ACE0}) are adequate for
illustrative purposes, while giving results essentially equivalent
to the `full' calculation.

\par
As shown by Eq.~(\ref{ACE1}), the peculiar property
of the observable $A_{\rm CE}$, as a function of
$z^*$, is that it is the same for {\it all} spin-1
exchanges, the SM $\gamma,Z$ and any
$Z^\prime$ model, regardless of the actual values
of the left- and right-handed coupling constants
to fermions, of the $Z^\prime$ mass $M_{Z^\prime}$
and, to a large extent, of the choice
of parton distribution functions. Deviations of
$A_{\rm CE}$ from the SM predictions can therefore
be attributed to spin-2 or spin-0
exchanges in (\ref{proc_DY}).

\par To assess the domains in which the spin-0 and spin-2 hypotheses can be
excluded as sources of a peak observed at $M=M_R$, while giving the same
number of events in the assigned interval $\Delta M$ in Eq.~(\ref{Eq:TotCr})
we start from the assumption that spin-1 is the `true' origin of the
resonance.  The level at which the two alternative hypotheses can be excluded
for all `allowed' values of their relevant model parameters, is determined by
the experimental data, from the prediction of spin-1 ($V$) exchange, from the
predicted spin-0 ($S$) and spin-2 ($G$) exchanges, respectively:
\begin{equation}
\label{Eq:Delta-ACE}
\Delta A_\text{CE}^S=A_\text{CE}^S-A_\text{CE}^V
\qquad {\rm and}\qquad
\Delta A_\text{CE}^G=A_\text{CE}^G-A_\text{CE}^V.
\end{equation}
Of course, the identification potential will depend
on the available statistics, i.e., from the number
of signal events collected in the assigned
$M$-interval,
in addition to the systematic uncertainties.
The latter are however expected to have a reduced
influence on $A_{\rm CE}$,
because it is a ratio of cross
sections.

%%%%%%%%%%%%%%%%%%%%%%%%%%%%
\begin{figure}[tbh] % Fig.2
%\vspace*{-2.0cm}
\centerline{ \hspace*{-0.0cm}
\includegraphics[width=7.5cm,angle=0]{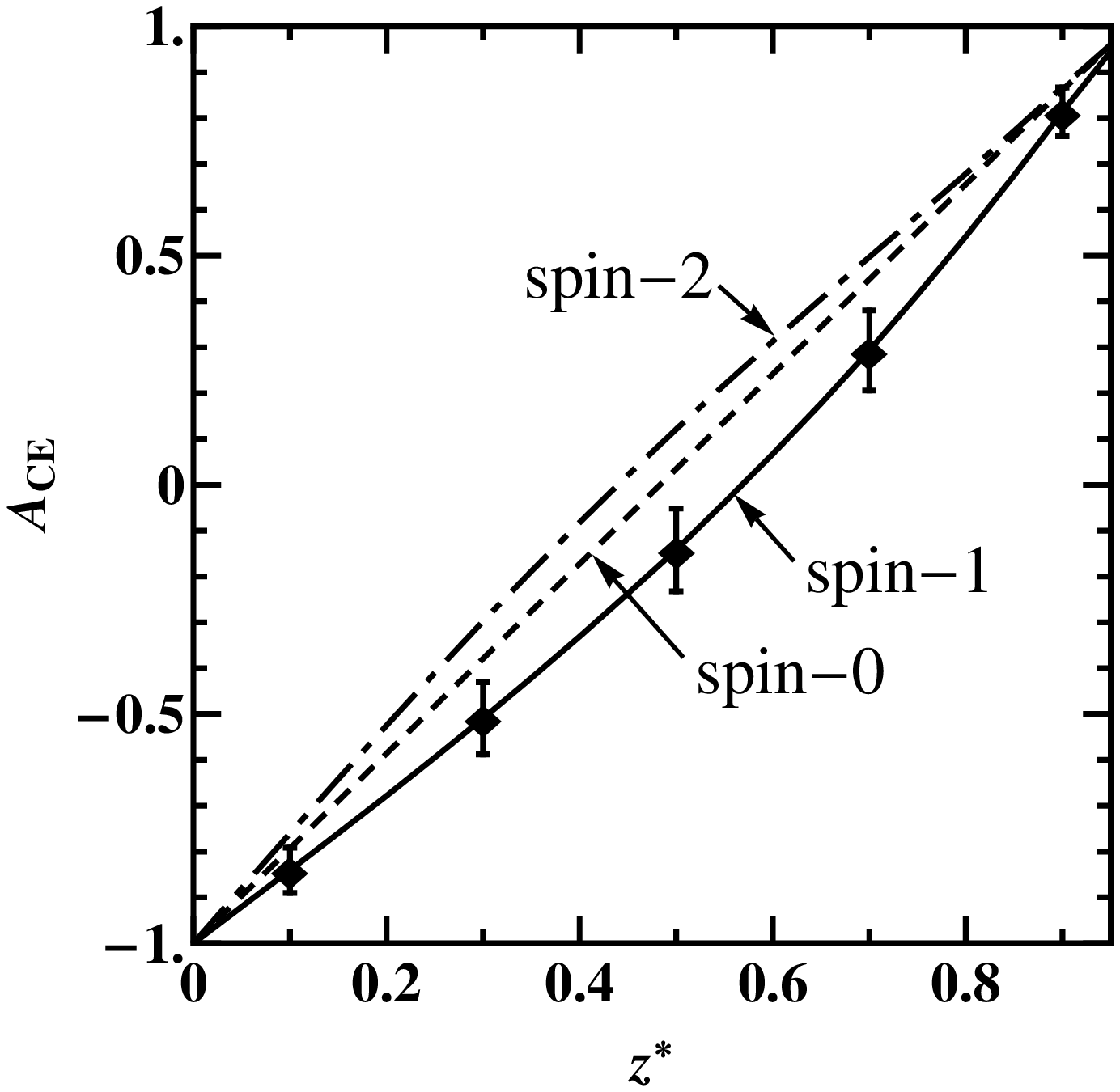}
\hspace*{0.4cm}
\includegraphics[width=7.5cm,angle=0]{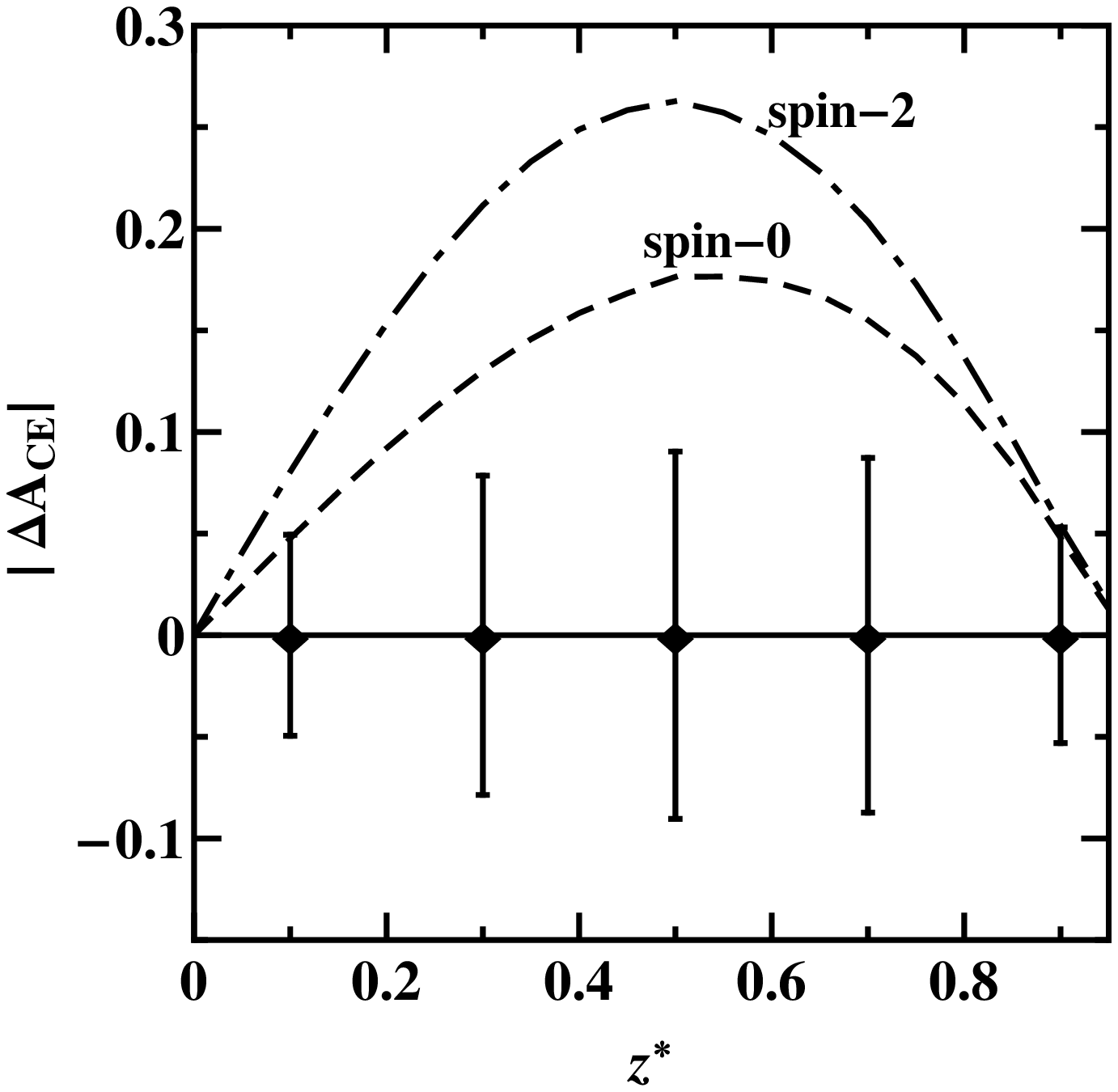}}
%\vspace*{-7.5cm}
\caption{\label{fig2}Left panel: $A_{\rm {CE}}$ vs.\ $z^{*}$ for
resonances $\tilde{\nu}$ (spin-0), $Z'$ (spin-1) and $G$ (spin-2)
with equal masses of 3 TeV. The error bars are the statistical
uncertainties at the 1-$\sigma$ level on $A_{\rm
CE}^{\text{spin-1}}$
for the $\psi$-model at $100~\text{fb}^{-1}$.
Right panel: Asymmetry deviations, $|\Delta
A_{\rm {CE}}|$, of the spin-0 and spin-2 hypotheses from the
spin-1 one, compared with the uncertainties on $A_{\rm
{CE}}^{\text{spin-1}}$. }
\end{figure}
%%%%%%%%%%%%%%%%%%%%%%%%%%

As an example in Fig.~\ref{fig2}, left panel, the center-edge asymmetry
$A_{\rm CE}$ is depicted as a function of $z^*$ for resonances with different
spins, the same mass $M_R=3$~TeV and the same number $N_S$ of signal events
under the peak. As anticipated, the calculations are performed using detector
cuts and, also, the SM background has been accounted for.  The deviations
(\ref{Eq:Delta-ACE}) are plotted in the right panel of the figure. The
vertical bars attached to the solid line represent, again as an example, the
1-$\sigma$ statistical uncertainty on the $A_{\rm CE}^V$ corresponding to the
$Z^\prime_\psi$-model with the assigned mass $M_{Z^\prime}$.  Fig.~\ref{fig2}
shows that the $Z^\prime_\psi$-model with mass $M_{Z^\prime}=3$~TeV can be
discriminated from the other spin-hypotheses at the 2-$\sigma$ level by means
of $A_\text{CE}$ at $z^*\simeq 0.5$.  
\par While $A_\text{CE}^V$ is independent of energy, $A_\text{CE}^G$ and
$A_\text{CE}^S$ are not. In the limit of little background,
$\epsilon_q^\text{SM}$ will be small, and $A_\text{CE}^S$ will only depend
weakly on the energy.  On the other hand, even in this limit, $A_\text{CE}^G$
will in general have a significant dependence on energy, via the relative
magnitudes of the fractions $\epsilon_q^G$ and $\epsilon_g^G$.  An exception
to this energy dependence is the region around $z^*\simeq 0.5$, where the
coefficient of $\epsilon_q^G$ vanishes. In this case, we have
\begin{equation}
A_{\rm CE}^G(z^*\simeq 0.5)>A_{\rm CE}^S(z^*\simeq 0.5). \label{DeltaA1-0}
\end{equation}
This property is of course reproduced in Fig.~\ref{fig2},
and allows to conclude that, in order to identify the
spin-1 $Z^\prime$ resonance, if one is able to
exclude the spin-0 hypothesis, the  spin-2 graviton of the RS
model will then automatically be excluded.

\par In order to determine the spin-1 signature space where the spin-0
hypothesis could be excluded against the spin-1 one, the deviation
(\ref{Eq:Delta-ACE}) should be compared with the statistical uncertainty on
$A_\text{CE}$ expressed in terms of the desired number ($k$) of standard
deviations ($k^2=3.84$ for 95\% C.L.). Notice that in practice $A_\text{CE}$
is almost unaffected by systematic uncertainty being a relative quantity.  We
have the condition
\begin{equation}  \label{Eq:StandDev}
\vert\Delta A_\text{CE}^S\vert=k\cdot\delta A_\text{CE}^V,
\end{equation}
where, taking into account that numerically $(A_\text{CE}^V)^2\ll
1$ at $z^*\simeq 0.5$,
\begin{equation} \label{Eq:stat}
\delta A_\text{CE}^V
=\sqrt{\frac{1-{(A_\text{CE}^V)}^2}{N_{\rm{min}}}} \approx
\sqrt{\frac{1} {N_{\rm{min}}}}.
\end{equation}
From Eqs.~(\ref{Eq:StandDev}) and (\ref{Eq:stat}), one therefore obtains
\begin{equation} \label{Eq:SPIN0-EXCL}
N_{\rm{min}}=N_{\rm{min}}^S
\approx\left(\frac{k}{\epsilon_q^V A_{\text{CE}}^V}\right)^2,
\end{equation}
using the fact that 
$\vert A_{\text{CE}}^S(z^*\simeq0.5)\vert$ is small compared to
$\vert A_{\text{CE}}^V(z^*\simeq0.5)\vert$, and
$\Delta A_\text{CE}^S=-\epsilon_q^V A_\text{CE}^V$
with $\epsilon_q^V=1-\epsilon_q^\text{SM}\simeq1$. 
From Eq.~(\ref{Eq:SPIN0-EXCL}) one can easily
evaluate  the minimal number of events required to exclude
the spin-0 hypothesis and, automatically, spin-2 as well, and in
this way to establish the spin-1. One finds using
Eq.~(\ref{Eq:SPIN0-EXCL}) for the
exclusion of the spin-0 resonance at 95\% C.L.,
$N_{\rm{min}}\simeq 110$. One should emphasize that $N_{\rm{min}}$
determined from Eq.~(\ref{Eq:SPIN0-EXCL}) is a
model-independent value, since $A_{\text{CE}}^V$ defined in
Eq.~(\ref{ACE1}) is independent of specific $Z'$ models, being
`universal' for all spin-1 intermediate states. Accordingly,
spin-1 of the discovered resonance can be established at 95\% C.L.
if resonance event samples $N_S$ at the level of
$N_{\rm{min}}$ or larger would be available.
\par
The behavior of $N_{\rm min}^{S}$ {\it vs.}\ $M_R$, as presented in
Fig.~\ref{fig1}, is derived from the full calculation including
detector cuts, using the general Eq.~(\ref{Eq:StandDev}). The
intersection of the curves describing $N_S$ against $M_R$ for
specific $Z'$ models and displayed in Fig.~\ref{fig1} with the
line of $N_{\rm min}$ determines the values of the $Z'$ masses
where the spin-1 hypothesis can be identified. One finds that for
$M_{Z'}\le 3$ TeV the spin of $Z'$ can be determined at 95\% C.L.
for all models under study, at 14~TeV, with $\Lumint=100$
fb$^{-1}$. 
For completeness we also display in Fig.~\ref{fig1}
$N_{\rm min}^{G}$ that lies below $N_{\rm min}^{S}$ as
anticipated.

In addition to the illustrative consideration above, one can
quantify the identification reach on the spin-1 hypothesis
performing a `conventional' $\chi^2$ analysis to obtain the
exclusion domains of the spin-2 and spin-0 hypotheses. In this
case, the $\chi^2$ function is defined as:
\begin{equation} \label{Eq:five-three}
\chi^2=\left[\frac{\Delta A_\text{CE}} {\delta
A_\text{CE}}\right]^2,
\end{equation}
with $\Delta A_{\rm CE}$ the deviations (\ref{Eq:Delta-ACE})
and $\delta A_{\rm CE}$ the statistical
uncertainty (a specific spin-1 $Z'$ model is taken as a the `true' one)
\begin{equation} \label{Eq:StatUncert}
\delta A_\text{CE}=\sqrt{\frac{1-{(A_\text{CE}^V)}^2} {\epsilon_l
{\cal L}_\text{int}\sigma(V_{ll})}}.
\end{equation}
%%%%%%%%%%%%%%%%%%%%%%%
\begin{figure}[tbh!] % Fig.3
\vspace*{-0.0cm} \centerline{ \hspace*{-2.0cm}
\includegraphics[width=12.0cm,angle=0]{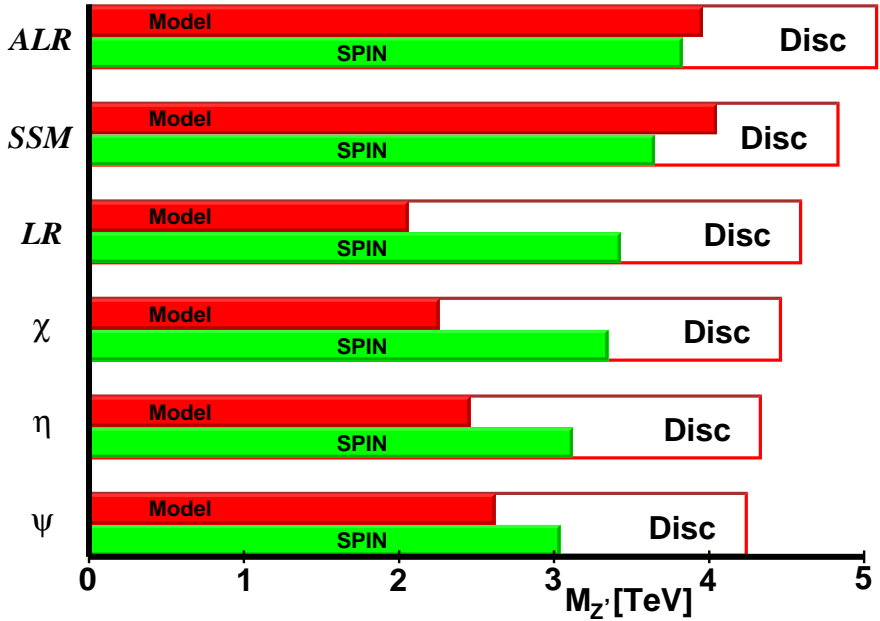}}
\vspace*{-0.0cm} \caption{\label{fig3} Discovery limits on
$M_{Z'}$ (5-$\sigma$ level) and $Z^\prime$-spin identification
reaches (95\% C.L.) for neutral gauge bosons of representative
models, using the lepton-pair production cross section
$\sigma\cdot B_l$ ($l=e, \mu$) and center-edge asymmetry
$A_\text{CE}$, respectively, at the LHC with integrated luminosity
of 100 fb$^{-1}$. Also, $Z^\prime$-model distinction reaches (95\%
C.L.) are obtained from the analysis of the leptonic event rates.
}
\end{figure}
%%%%%%%%%%%%%%%%%%%%%%%%%%%%%%%%%%%%%%%
\par Like before, the spin-1 $V$ model can be assumed to be the `true' one,
and the (95\% C.L.) exclusion domains of spin-0 and spin-2 can be determined.
Here, we combine the channels $l=e,\mu$. The 95\% C.L. identification reach of
the spin-1 hypothesis in the signature space then results from the domain
complementary to the combination of the spin-0 and spin-2 exclusion
domains. The results of this numerical analysis are represented in the
signature space $(M_{Z'}, N_S)$ in Fig.~\ref{fig1}. In fact, the spin-0
exclusion is more restrictive than that for spin-2, as discussed
above. Fig.~\ref{fig3} gives the `translation' of the discovery reach on $Z'$
models as well as identification reach on $Z'$ spin presented in
Fig.~\ref{fig1}, in the form of a histogram. As one can see from 
Fig.~\ref{fig3}, the spin-1 identification (or, actually, the spin-0 
and spin-2 exclusion) can be obtained up to $Z^\prime$ mass of the order 
of 2.5--3.5 TeV, depending on the specific models. The model dependence 
of the spin identification reach is due to the difference in statistics,
stemming from the different cross sections associated with these models. 

\par It might be useful to conclude this section by emphasizing that 
the basic observable $A_{\rm CE}$ of Eq.~(\ref{Eq:ace}) only uses 
the $z=\cos\theta_{\rm c.m.}$ even part of the angular differential 
distribution, similar to the total cross section. The `center' and 
`edge' angular integration regions are symmetric around $z=0$ and 
include events with both signs of $z$. This might mitigate the impact 
of the ambiguity, at the $pp$ colliders, in the experimental 
determination of the sign of $z$ that can affect observables sensitive 
to the $z$-odd part of the angular distribution.     

%%%%%%%%%%%%%%%%%%%%%%%%%%%%%%%%%%%%%%%%%%
\section{Differentiating $Z'$ models}
\label{sect:LHCdiff}
\setcounter{equation}{0}
%%%%%%%%%%%%%%%%%%%%%%%%%%%%%%%%%%%%%

Once the spin-1 character of a $Z^\prime$ peak at $M=M_R\equiv M_{Z^\prime}$
has been verified by the exclusion of the spin-0 and spin-2 hypotheses, one
can attempt the more ambitious task of identifying the resonance with one of
the $Z^\prime$ models by means of the measured production cross section
$\sigma\cdot B_l$ or, equivalently, of the peak event rate $N_S$. As 
anticipated in Subsec.~\ref{subsect:statistics}, $B_l$ will be 
assessed under the simplifying assumption of $Z^\prime$ decays to SM 
fermions only. Results from other observables, such as 
$\sigma\cdot\Gamma_{Z^\prime}$, $A_{\rm FB}$, etc., have been 
qualitatively summarized in 
Sec.~\ref{sect:introduction}.\footnote{Actually, a precise measurement 
of the ratio $\Gamma_{Z^\prime}/M_{Z^\prime}$, if feasible, might also 
represent a discrimination criterion among classes of $Z^\prime$ models 
by itself.} 

%%%%%%%%%%%%%%%%%%%%%%%%%%%%
\begin{figure}[tbh] % Fig.4
%\vspace*{-2.0cm}
\centerline{ \hspace*{-0.0cm}
\includegraphics[width=7.5cm,angle=0]{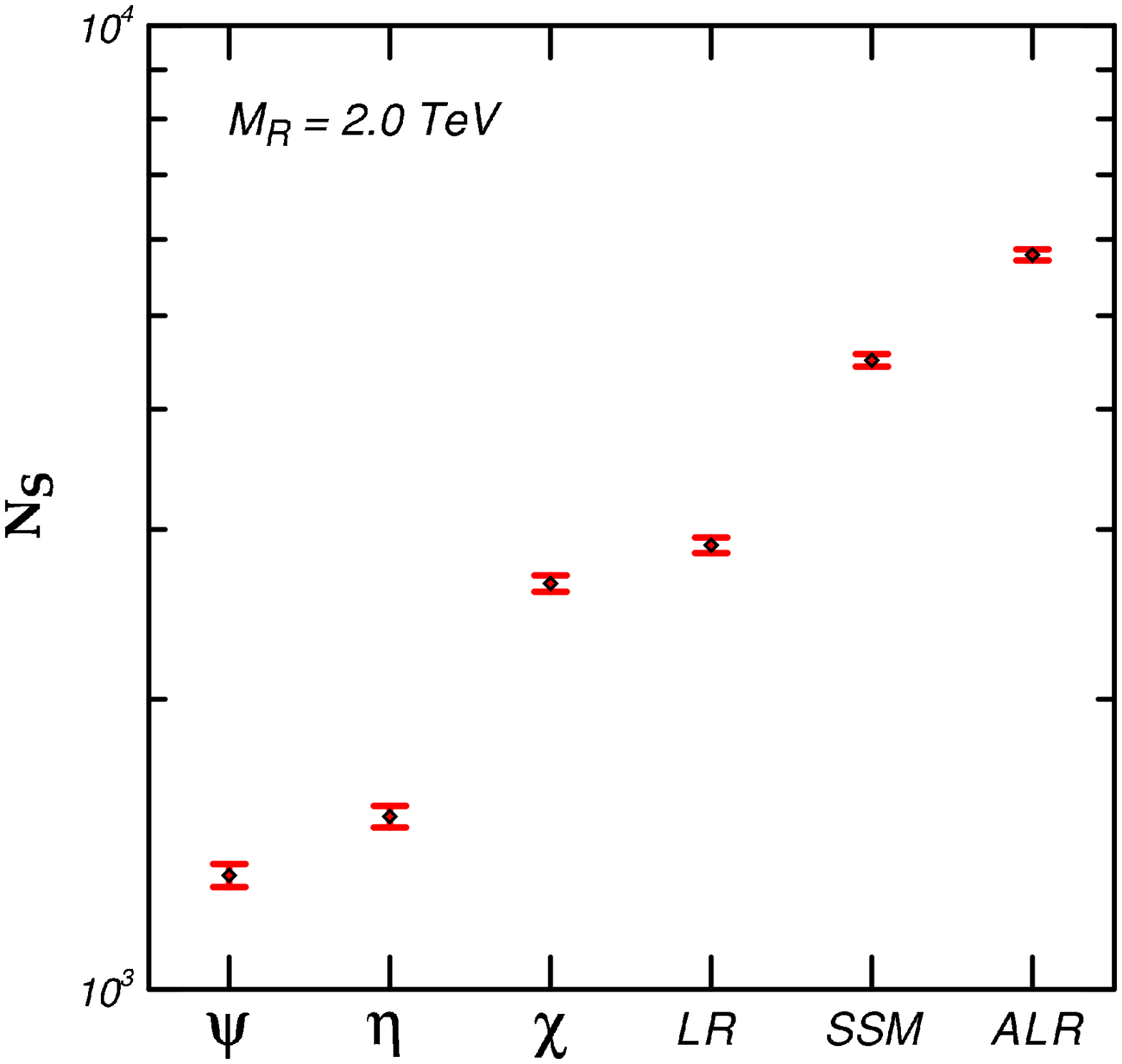}
\hspace*{0.4cm}
\includegraphics[width=7.5cm,angle=0]{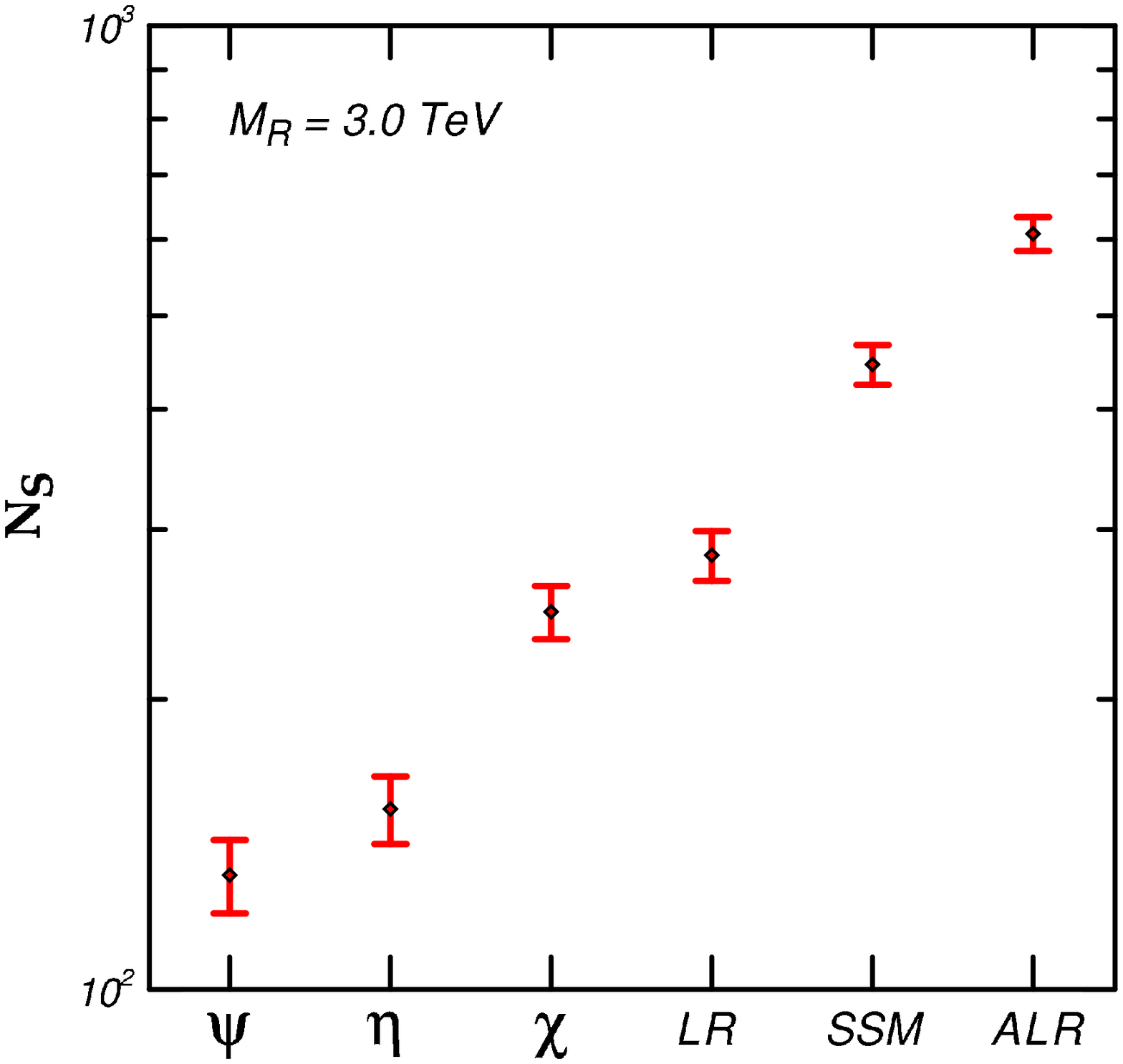}}
%\vspace*{-7.5cm}
\caption{\label{fig4} Resonance event rates obtained for the
reference models at $M_{Z'}=2$ TeV (left panel) and 3 TeV (right
panel) are displayed. The error bars show the 
1-$\sigma$ statistical uncertainty at 100 fb$^{-1}$ of 
integrated luminosity.}
\end{figure}
%%%%%%%%%%%%%%%%%%%%%%%%%%

\par
One can see from Fig.~\ref{fig4}, representing $Z^\prime$ signal 
event rates for two specific values of the $Z^\prime$ mass, that 
at $M_{Z'}=2$ TeV, an integrated luminosity of 100 fb$^{-1}$ 
will be sufficient to distinguish all considered models pairwise, 
whereas at $M_{Z'}=3$ TeV, one is unable to distinguish 
neither between $Z_\psi$ vs.\ $Z_{\eta}$ nor between 
$Z_\chi$ vs.\ $Z_{\rm LR}$.

\par To perform this discrimination, we make the hypothesis 
that one of the considered models ($Z^\prime_{\rm SSM}$, 
$Z^\prime_{\psi}$, $Z^\prime_{\eta}$, $Z^\prime_{\chi}$, 
$Z^\prime_{\rm LR}$, or $Z^\prime_{\rm ALR}$) is the `true' one, 
compatible with the measured cross section, and test this 
assumption against the remaining five models that in general 
can predict, for the same $M_{Z^\prime}$, a different value 
of the cross section but within the uncertainty band of the 
former, hence not distinguishable from it. As a simple criterion, 
one can define a `separation', in peak
event rates $N_S$, between the `true model' and the others, and
then associate the foreseeable identification reach on the chosen
`true' model to the maximum value of $M_{Z^\prime}$ for which all five 
such separations are larger than an amount specified by a desired
confidence level. Finally, one can iterate this numerical
procedure, in turn, for all six considered $Z^\prime$ models.
\par
\begin{figure}[tbh] % Fig.5
\vspace*{0.0cm} \centerline{ \hspace*{-2.0cm}
\includegraphics[width=11.0cm,angle=0]{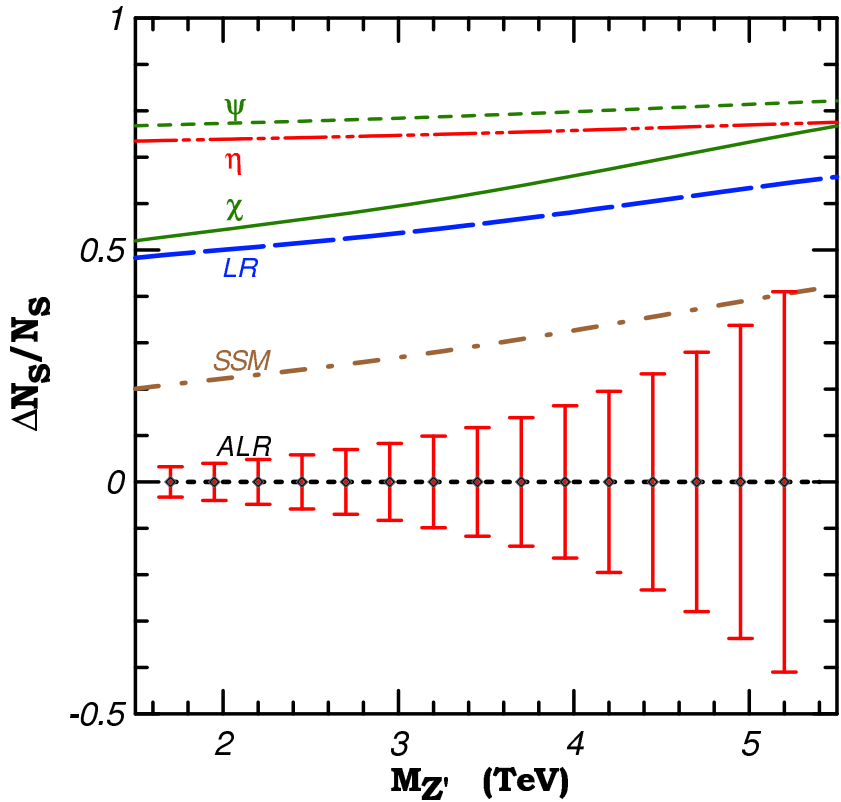}}
\vspace*{0.cm} 
\caption{\label{fig5} Absolute value of relative
deviation of  the number of events, Eq.~(\ref{delta_N}), from the
$Z'_{\rm ALR}$ predictions shown in Fig.~1 as a function of
$M_{Z^\prime}$ for neutral gauge bosons of representative models.
The error bars are the uncertainties at the 1-$\sigma$
level on $\Delta N_S/N_S$ for the ALR model, $\Lumint=100\, {\rm bf}^{-1}$.}
\end{figure}

For definiteness, we work out explicitly the example 
of the identification reach on the $Z^\prime_{\rm ALR}$ model.
We introduce the relative deviations of the event rates predicted
by this model, at the generic $M=M_{Z^\prime}$, from those predicted
by the other $Z^\prime$ models:
\begin{equation}
\frac{\Delta N_S}{N_S} = \frac{N_S(Z')-N_S(Z'_{\rm
ALR})}{N_S(Z'_{\rm ALR})}. \label{delta_N}
\end{equation}
Figure~\ref{fig5} shows the relative deviations (\ref{delta_N}) as
functions of $M_{Z'}$. Vertical bars represent the 1-$\sigma$ combination 
of the statistical uncertainty predicted by the ALR model at integrated LHC
luminosity ${\cal L}_{\rm int}=100\, {\rm fb}^{-1}$, with the major systematic 
uncertainty for the total cross section $\sigma\cdot B_l$, represented by the 
uncertainties on the parton distribution functions (PDF). These are 
calculated using the CTEQ6.5 NLO PDF sets~\cite{Pumplin:2002vw}. 
%%%%%%%%%%%%%%%%%%%%%%%%%%%%%%%%%%%%%%%%%%%%%%%%%%%%%%%%%%%%%%%%%%%%%%%%%%
%%%%%%%%%%%%%%%%%%%%%%%%%%%%%%%%%%%%%%%%%%%%%%%%%%%%%%%%%%%%%%%%%%%%%%%%%%

Corresponding to the definition (\ref{delta_N}), one can introduce
a ${\chi}^2$ function
\begin{equation}
\chi^2=\left(\frac{\Delta {N_S}}{\delta N_{S}}\right)^2
\label{chi2}
\end{equation}
with ${\delta}N_{S}$ the corresponding experimental uncertainty,
which includes both the statistical and systematic errors combined
in quadrature, the former being determined by the $Z'_{\rm ALR}$ model
prediction of the event rate. 
It turns out that for $M \lsim 4$ TeV the systematic uncertainty is 
larger than the statistical one, they cross over at a value around 15\%.
This systematic uncertainty has 
an effect comparable to that of the statistical uncertainty 
for $Z'$ resonances with masses larger than 3 TeV \cite{Petriello:2008zr}.
\par
On the basis of such ${\chi}^2$ we can determine the maximum value
of $M_{Z^\prime}$ (hence the identification reach) for which the
$Z'$ models, with $Z'\neq Z'_{\rm ALR}$, can be excluded once the
ALR model has been assumed to be the `true' one. This value must
satisfy the conditions $\chi^2>\chi^2_{\rm C.L.}$ for all
$Z^\prime$s, where $\chi^2_{\rm C.L.}$ determines the chosen
confidence level. The
results of this procedure, applied in turn to all six $Z^\prime$
models, are reported in Fig.~\ref{fig3}, together with the
discovery and the spin-1 identification reaches, and indicate that 
all models under considerations might be distinguishable up to 
$M_Z^\prime$ of the order of 2.1~TeV.

\par Of course, these results rely numerically on the 
assumption about $Z^\prime$ decay stated at the beginning. On 
the other hand, neither of the $Z^\prime$ curves 
in Fig.~\ref{fig1} intersects with any other, so that the 
simple (and directly measurable) total cross section might also 
be considered a natural discriminator among models.         

%%%%%%%%%%%%%%%%%%%%%%%%%%%%%%%%%%%%%%%%%%%%%%%%%%%%
\section{The reduced-energy, low-luminosity case}
\label{sect:low-low}
\setcounter{equation}{0}
%%%%%%%%%%%%%%%%%%%%%%%%%%%%%%%%%%%%%%%%%%%%%%%%%%%

It may take quite some time before the experiments will be able 
to collect $100~{\rm fb}^{-1}$ of data at 14~TeV. In the present 
section, we indicate how the spin-identification reach depends on the 
integrated luminosity, at two different energies, 10 and 14~TeV. 
The former is expected to be the energy initially available at the LHC. 

\par
\begin{figure}[tbh!] % Fig.6
\vspace*{-0.0cm} \centerline{ \hspace*{-2.0cm}
\includegraphics[width=10.0cm,angle=0]{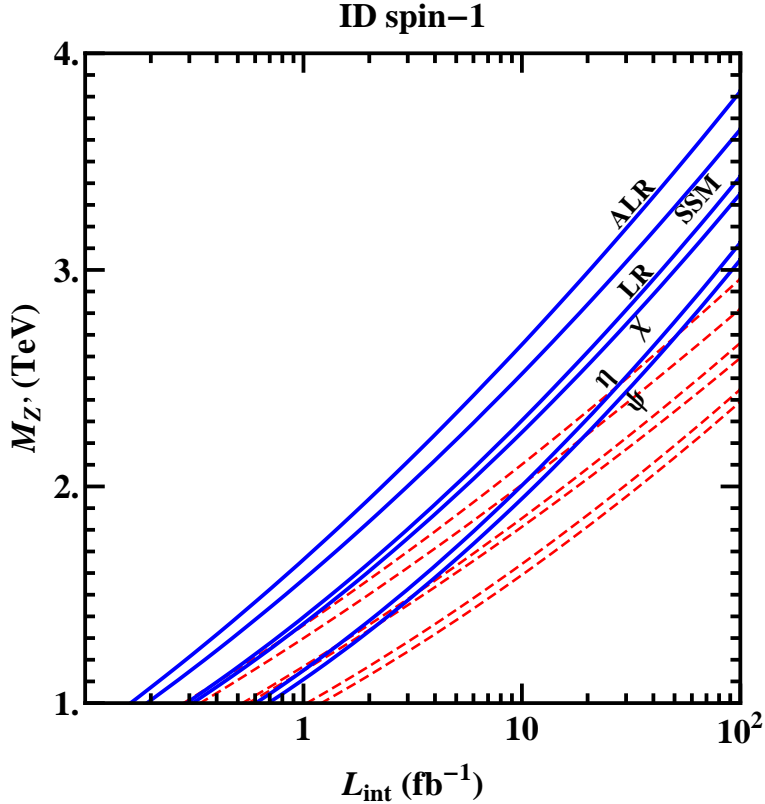}}
\vspace*{-0.0cm} \caption{\label{fig6} Spin-determination reach as a function
of integrated luminosity, for 10~TeV (dashed) and 14~TeV (solid).  }
\end{figure}

 Of course, in general, a detailed assessment of the spin-identification
dependence vs.\ the total energy and the $Z^\prime$ mass depends on the, in
some cases complicated behaviour of the individual parton distribution
function, in addition to the applied cuts. Nevertheless, for a simplified
estimate leading to an explicit, parametric, expression of such energy
dependence, in a very rough sense one can trade energy for luminosity. Indeed,
according to Sec.~\ref{sect:spin1-ID}, the procedure of $Z^\prime$ spin-1
determination using $A_{\rm CE}$ basically consists in excluding the spin-0
case, since spin-2 is then automatically excluded. The spin-0 cross section is
determined by the parton cross section together with the overlap of the $d$
and $\bar d$ parton distribution functions in the proton,
\begin{equation}
I_{d\bar d}(s,M)=\int \dd y f_d(\xi_1,M)f_{\bar d}(\xi_2,M).
\label{Eq:overlap}
\end{equation}
The spin-1 cross section also depends on the corresponding overlap of the $u$
and $\bar u$ parton distribution functions, $I_{u\bar u}(s,M)$. In the
approximation that these two overlap integrals have the same dependence on
$E=\sqrt s$, and in the narrow-width approximation, all these cross sections
would have the same energy dependence, but they would differ by constant
ratios.  Thus, the spin-identification reach would be propoprtional to a
unique function of energy, or equivalently, integrated luminosity, the same
function for all $Z^\prime$ models. This reach would be given by the number of
events, proportional to the integrated luminosity, the parton cross section
$\hat\sigma$, and the overlap integral:
\begin{equation}
N\sim\Lumint \hat\sigma I(s,M).
\end{equation}
The spin identification reach basically requires a certain number of events
(around 110 at the higher values of $M_{Z^\prime}$), i.e., the scaling with
energy and luminosity is determined by keeping the above expression fixed as
the luminosity or energy is changed.  In the exponential approximation to the
overlap integral, given by Eq.~(3.16) of Leike \cite{Hewett:1988xc} (valid for
a wide range of $E/M_{Z^\prime}$), this becomes
\begin{equation}
N\sim\Lumint \frac{C_1}{E^2}\exp\left[-\frac{C_2 M_{Z^\prime}}{E}\right].
\end{equation}
Neglecting the energy and mass dependences of $C_1$ and $C_2$, one finds a
reach in $M_{Z^\prime}$ that grows linearly with $\log\Lumint$, with a slope
proportional to the beam energy, features which are qualitatively reflected in
Fig.~\ref{fig6}, which displays the spin-1 identification reach vs.\ the
$Z^\prime$ mass for the various models at the two energies mentioned
above. Experimental cuts have been taken the same as detailed in
Sec.~\ref{sect:models} for both cases.

%%%%%%%%%%%%%%%%%%%%%%%
%%%%%%%%%%%%%%%%%%%%%%%%%%%%%%%%%%%%%%%

\par Figure~\ref{fig6} speaks for itself. It shows that at 10~TeV, and
depending on the model, with a considerable fraction of $1~\text{fb}^{-1}$ one
could in principle identify the spin-1 of a hypothetical $Z^\prime$ with mass
in the range 1--1.5~TeV. This range is expected to be in the reach of the
Tevatron, for the planned luminosity increase from 2.5 up to 9 ${\rm fb}^{-1}$
\cite{Tev:2007sb}. However, the spin identification would be a unique feature
of the LHC.
%%%%%%%%%%%%%%%%%%%%%%%%%%%%%%%%%%%%%%%%%%%%%%%%%%%%%%%%%%%%%%%%%%%%%%%%%%
\section{Summary}
\label{sect:Concl}
\setcounter{equation}{0}
%%%%%%%%%%%%%%%%%%%%%%%%%%%%%%%%%%%%%%%%%%%%%%%%%%%%%%%%%%%%%%%%%%%%%%%%%%

We can summarize the main part of this paper, relevant to the $Z^\prime$
identification at 14~TeV and $100\, {\rm fb}^{-1}$ luminosity, as follows: if
new heavy resonance peaks will be discovered in the dilepton mass
distributions for process (\ref{proc_DY}), a $Z'$ can be observed up to
$M_{Z'}\approx 4-5$ TeV. The statistical significance of measurements of the
evenly-integrated (in $\cos\theta_{\rm c.m.}$) asymmetry $A_{\rm CE}$ will
allow to establish the spin-1 (or, to exclude the spin-0 and spin-2) of a
heavy $Z'$ gauge boson for $M_{Z'}\leq 3.0-3.8$ TeV, at 95\% C.L.  We also
assess the mass limits on $Z'$ for which the studied $Z'$ models can be
distinguished, besides their common spin-1, in pairwise comparisons with each
other.  By a simple criterion based on the expected statistics, we find that
with 100 $\text{fb}^{-1}$ of integrated luminosity one should be able to
distinguish among the six $Z'$ models up to $M_{Z'}\simeq 2.1$ TeV (95\%
C.L.).

\goodbreak
\vspace{0.5cm} 
\leftline{\bf Acknowledgements}
\par\noindent This research has been partially supported by the Abdus Salam
ICTP and the Belarusian Republican Foundation for Fundamental
Research. AAP also acknowledges the support of MiUR (Italian
Ministry of University and Research) and of Trieste University.
The work of PO has been supported by The Research Council of
Norway, and that of NP by funds of MiUR.

%%%%%%%%%%%%%%%%%%%%%%%%%%%%%%%%%%%%%%%%%%%%%%%%%%%%%%%%%%%%%%%%%%%%%%%%%

\end{document}